\documentclass[final,1p,times]{elsarticle} 

\usepackage{graphicx}
\usepackage{amssymb} 
\usepackage{amsthm} 
\usepackage{lineno}

\journal{Nuclear Physics A} 

\begin{document}

\begin{frontmatter} 

\title{Anisotropic flow of identified particles in Pb--Pb collisions at $\sqrt{s_{\rm NN}} = {\rm 2.76~TeV}$ measured with ALICE at the LHC}

\author{Francesco Noferini (for the ALICE\fnref{col1} Collaboration)}
\fntext[col1] {A list of members of the ALICE Collaboration and acknowledgements can be found at the end of this issue.}
\address{Museo storico della fisica e centro studi e ricerche ``E. Fermi'', Rome (Italy) and INFN, Bologna (Italy)}


\begin{abstract} 
We report on the elliptic and triangular flow measurements for a
number of hadrons including charged pions, and (anti-)protons, as well as
those with strangeness content: kaons,
${\rm \phi}$-meson, ${\rm K^0_s}$, ${\rm \Lambda/\bar{\Lambda}}$, ${\rm \Xi}$, and ${\rm \Omega}$.
The results reported cover mid--rapidity, $|\eta|<0.8$, and a wide,
$0.2 < p_{\rm T} < 16$ GeV/$c$, transverse momentum range for Pb--Pb
collisions at $\sqrt{s_{\rm NN}}$ = 2.76 TeV recorded by ALICE at the LHC.
The mass splitting and the scaling properties of the elliptic flow with the number of constituent quarks and the particle transverse mass are studied as a function of collision centrality. The results are compared to RHIC measurements and to hydrodynamic model predictions. 
\end{abstract} 

\end{frontmatter} 

\vspace{-0.3cm}
\section{Introduction}
\vspace{-0.3cm}
Anisotropic flow \cite{bib:v2} is an important observable to probe the nature of matter
produced in heavy-ion collisions and in general
to study collective effects among produced particles.
It is described by the coefficients in the Fourier expansion of the azimuthal
particle distribution with respect to the symmetry plane.
Anisotropic flow of identified hadrons is sensitive to the partonic degrees
of freedom at the early times of the heavy-ion collision evolution \cite{bib:v2phenom}.
It is sensitive to the properties of the deconfined QCD matter created
in a collision and details of the hadronization mechanism.
Herein, we report on the elliptic, $v_2$, and triangular, $v_3$, flow of identified particles
measured by the ALICE Collaboration for Pb--Pb collision at $\sqrt{s_{\rm NN}} = 2.76~{\rm TeV}$.
Results are compared to that at RHIC top energy.
The scaling properties of $v_2$ with the number of constituent quarks, $n_q$,
and kinetic energy, $KE_{\rm T}$ are studied.
\vspace{-0.3cm}
\section{Analysis details}
\vspace{-0.3cm}
A sample of about 10 million minimum bias Pb--Pb collisions
at $\sqrt{s_{\rm NN}} = 2.76~{\rm TeV}$ collected by ALICE in 2010 was used for the analysis.
The collision centrality is determined using the forward VZERO scintillator
arrays.
Particle tracking is done using the Time Projection Chamber
(TPC) and the Inner Tracking System (ITS) with full azimuth coverage for
$|\eta| < 0.8$.
Identification of pions, kaons and (anti-)protons is performed with a combination of
the time-of-flight and energy loss measured by the ALICE Time Of Flight (TOF) and TPC detectors.
${\rm K^0_s}$, ${\rm \Lambda/\bar{\Lambda}}$, ${\rm \Xi}$, and ${\rm \Omega}$ particles were reconstructed via their weak decay channels
(${\rm K^0_s \rightarrow \pi\pi}$, ${\rm \Lambda(\bar{\Lambda}) \rightarrow p(\bar{p})\pi}$, ${\rm \Xi \rightarrow \Lambda\pi}$, and ${\rm \Omega \rightarrow \Lambda K}$) using topological cuts to
suppress combinatorial background.
In case of ${\rm \phi}$ meson decay via KK channel,
the combinatorial background was suppressed by identifying kaons in the
TPC and TOF.
Anisotropic flow was measured using the scalar product and event plane methods
with large rapidity separation (1 or 2 units) between correlated particles \cite{bib:alice}.
\vspace{-0.3cm}
\section{Results}
\begin{figure}[htbp]
\begin{center}
 \includegraphics[width=0.49\textwidth]{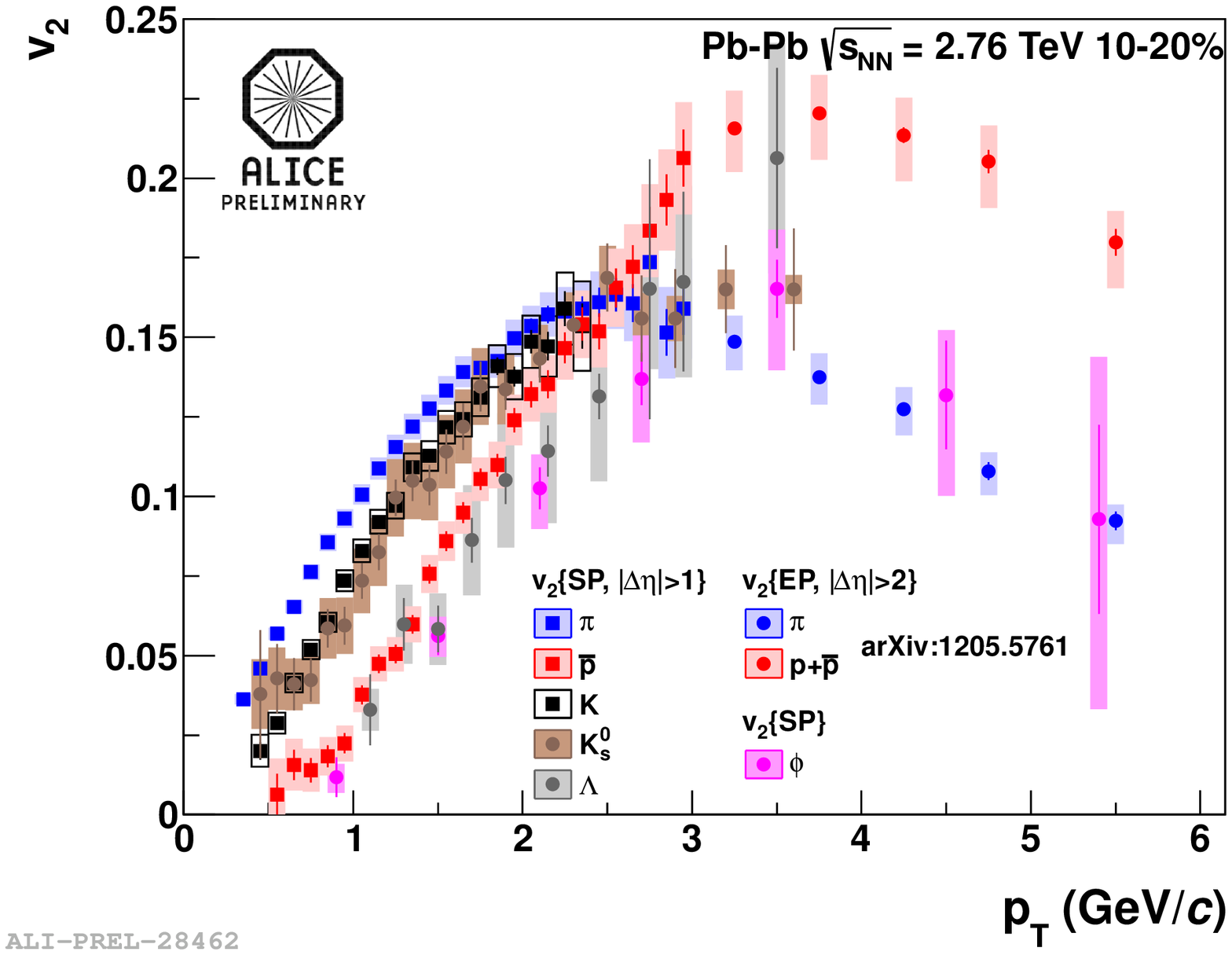}
 \includegraphics[width=0.49\textwidth]{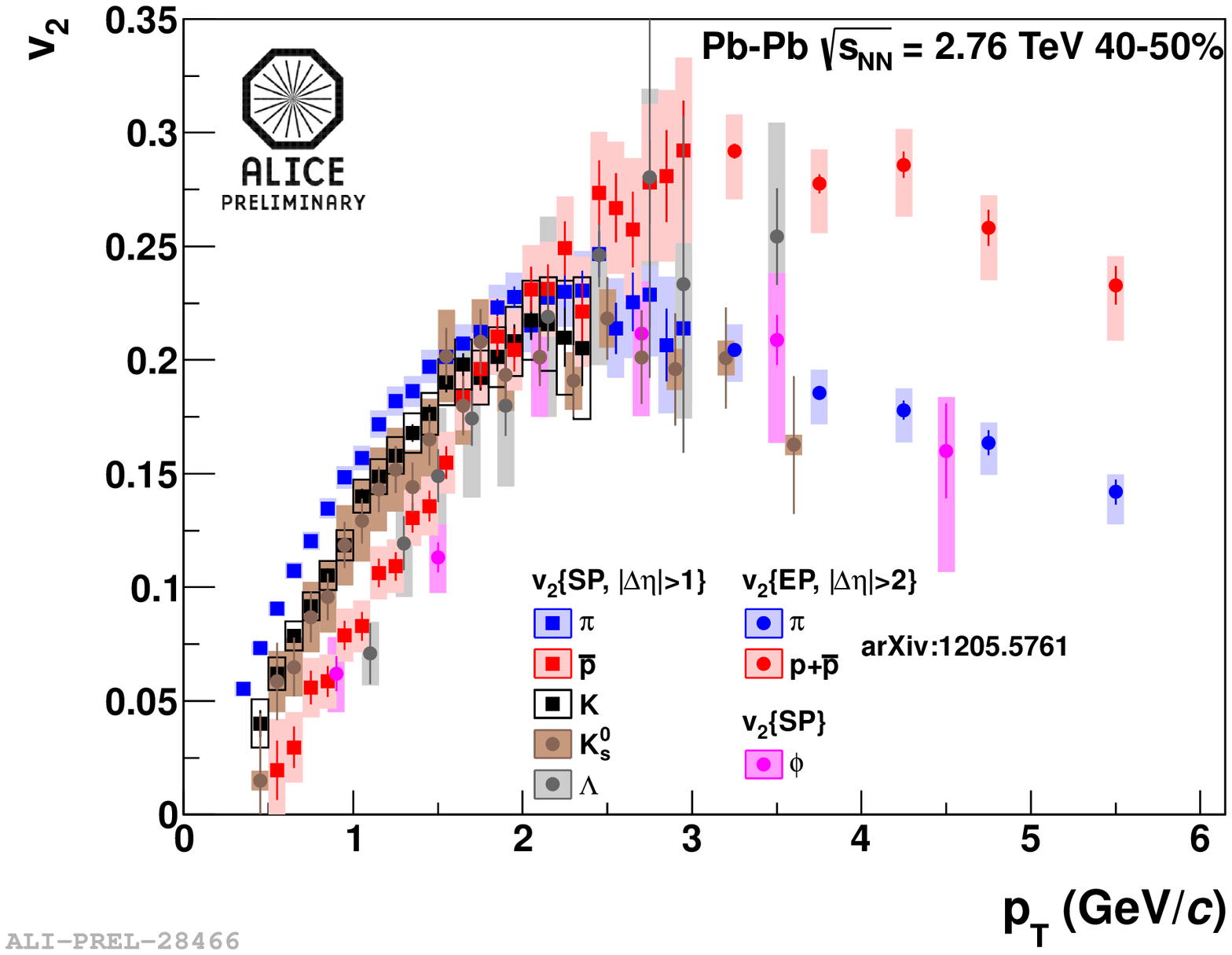}
\end{center}
\vspace{-0.8cm}
\caption{Identified particle $v_2(p_{\rm T})$  for  $p_{\rm T} <$ 6 GeV/$c$ measured by ALICE for 10-20\% (left) and 40-50\% (right) centrality classes.}
\label{fig:v2}
\end{figure}

\vspace{-0.3cm}
In Fig.~\ref{fig:v2} elliptic flow is presented for ${\rm \pi}$, K, p(${\rm \bar{p}}$), ${\rm K_s^0}$, ${\rm \Lambda(\bar{\Lambda})}$ and ${\rm \phi}$ for 10-20\% and 40-50\% centrality classes.
A clear mass ordering is seen for all species below $p_{\rm T} \sim 2.5~{\rm GeV/}c$,
where the ordering changes to show a clear baryon meson
difference.
At low transverse momenta, $p_{\rm T}$, the ${\rm \phi}$-meson elliptic flow is similar to that of (anti-)proton.
This can be explained by a similar effect of radial flow for particles with similar mass.
At higher momenta, ${\rm \phi}$-meson $v_2$ is closer to that of pion (also meson),
which one would expect  for a coalescence mechanism of hadron production in the intermediate $p_{\rm T}$ region.
A similar mass ordering is observed for multi-strange baryons ${\rm \Xi}$ and ${\rm \Omega}$ shown in Fig. 2(left).
Viscous hydrodynamic model calculations reproduce the
mass ordering of $v_2$ at low transverse momentum for all species.
A better agreement is observed for the peripheral collisions.
For central collisions agreement with proton $v_2$ can be improved by
adding the phase of hadronic rescattering with UrQMD model as shown in \cite{bib:hydro}.
Figure \ref{fig:v2rhic} shows a comparison of ALICE results to $v_2$ measurements
for pions, kaons and proton by the PHENIX \cite{bib:phenix} and the ${\rm \phi}$-meson by the STAR \cite{bib:star} Collaborations.
\begin{figure}[htbp]
\begin{center}
\includegraphics[width=0.49\textwidth]{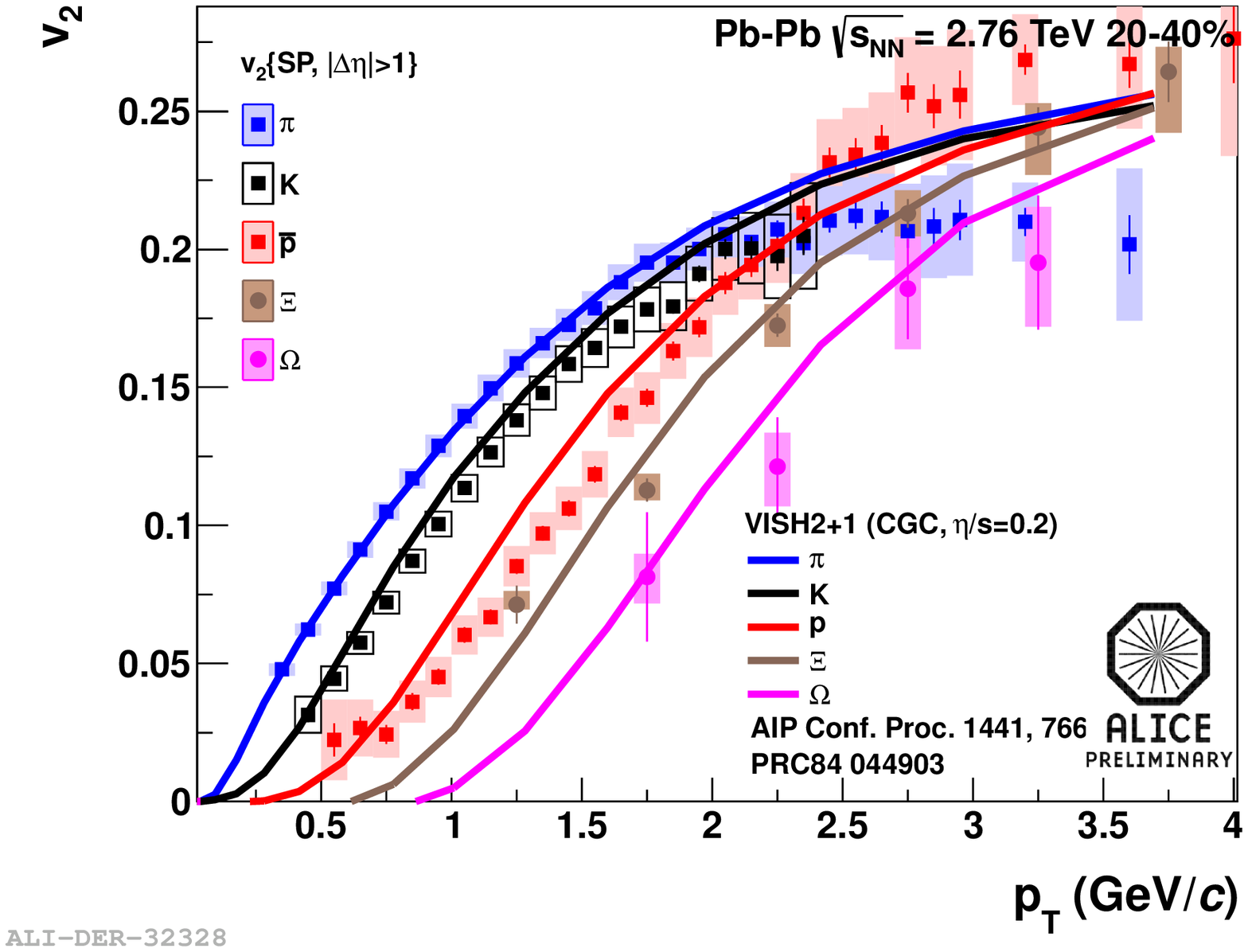}  \includegraphics[width=0.49\textwidth]{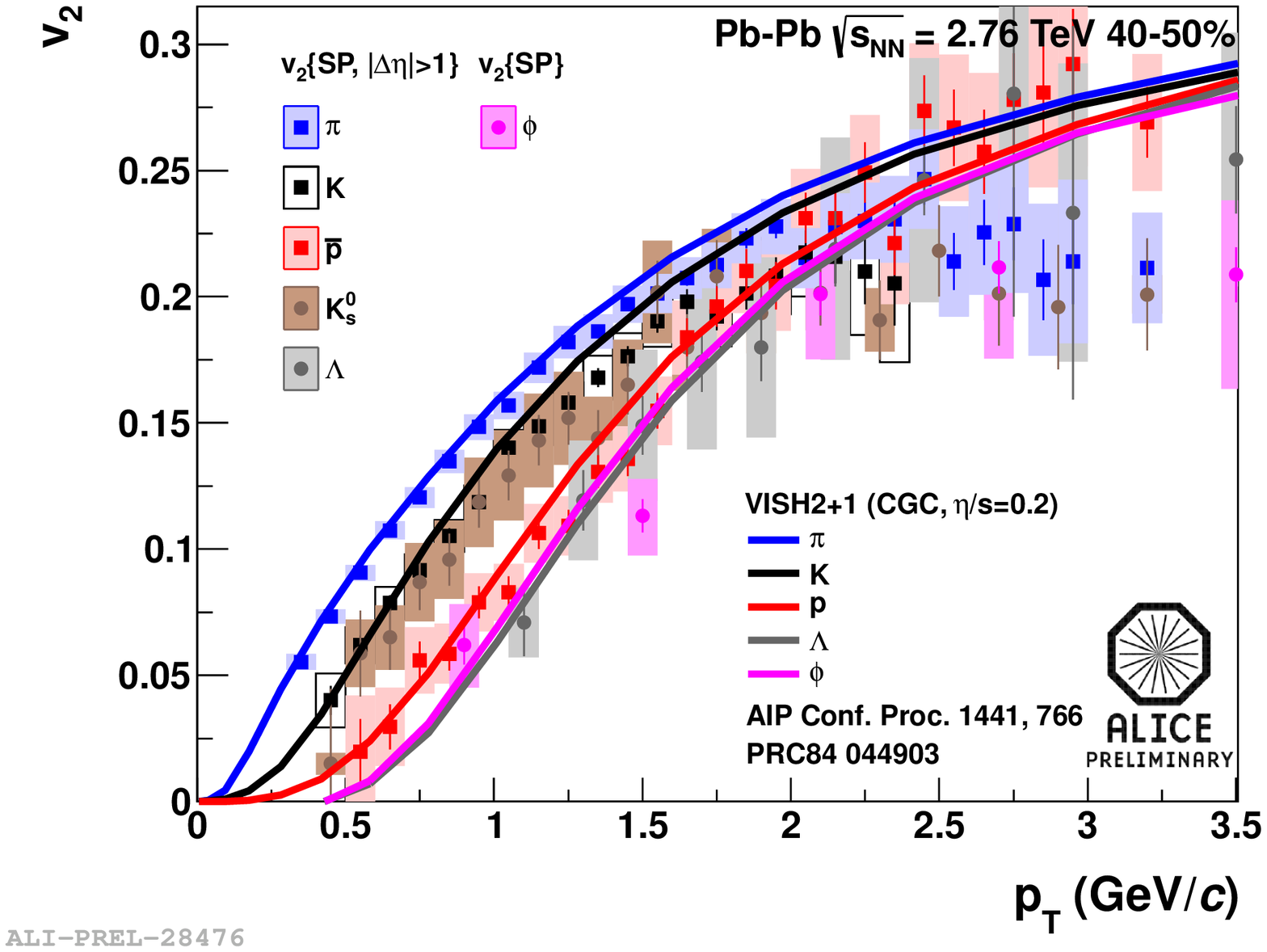}
\end{center}
\vspace{-0.8cm}
\caption{Identified particle $v_2(p_{\rm T})$ measured by ALICE vs. viscous hydrodynamic model calculations \cite{bib:hydro} for 20-40\% centrality focusing on $\Xi$ and $\Omega$ flow (left) and for 40-50\% centrality focusing on $\phi$-meson flow (right).}
\label{fig:v2hydro}
\end{figure}

\begin{figure}[htbp]
\begin{center}
 \includegraphics[width=0.49\textwidth]{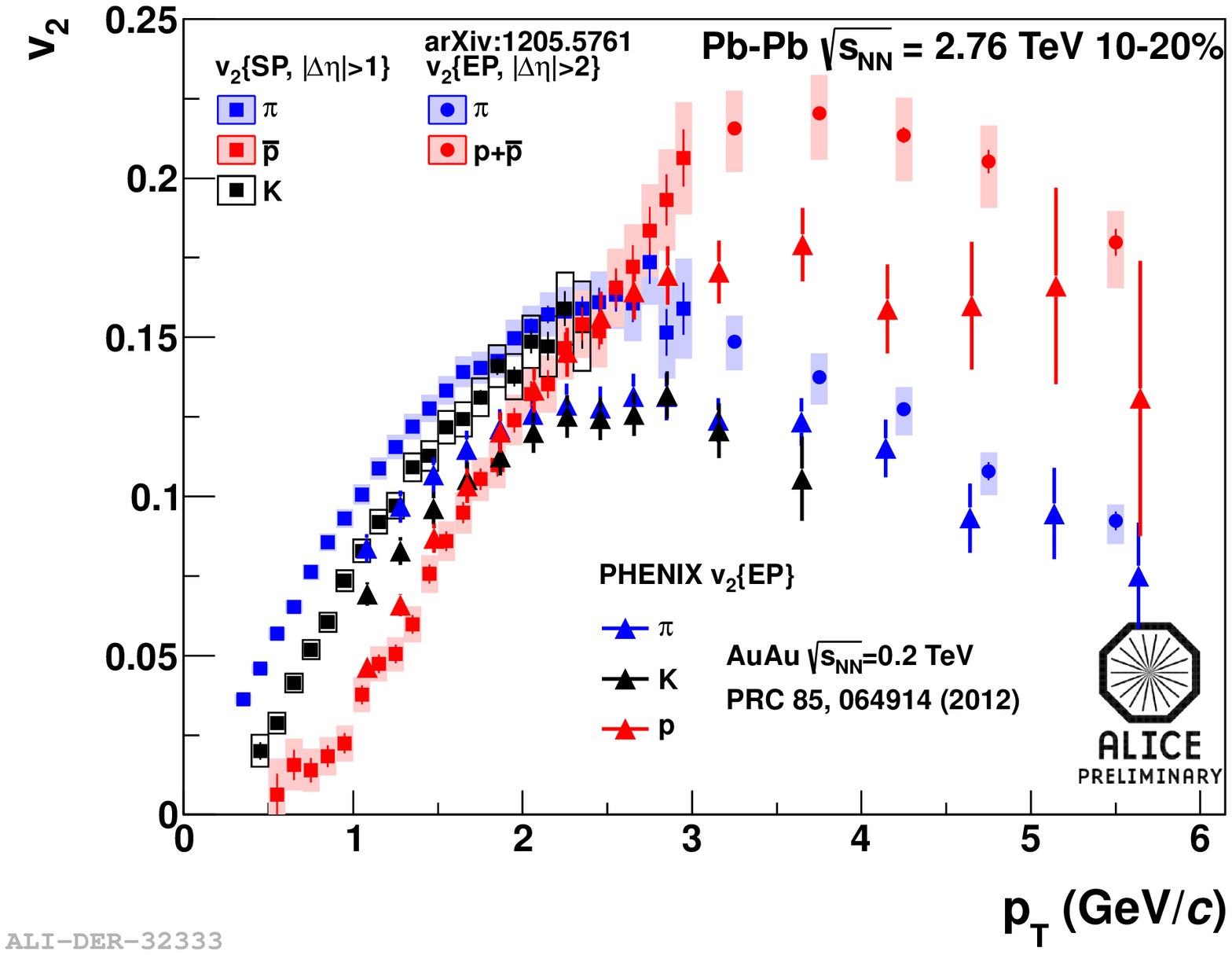}
 \includegraphics[width=0.49\textwidth]{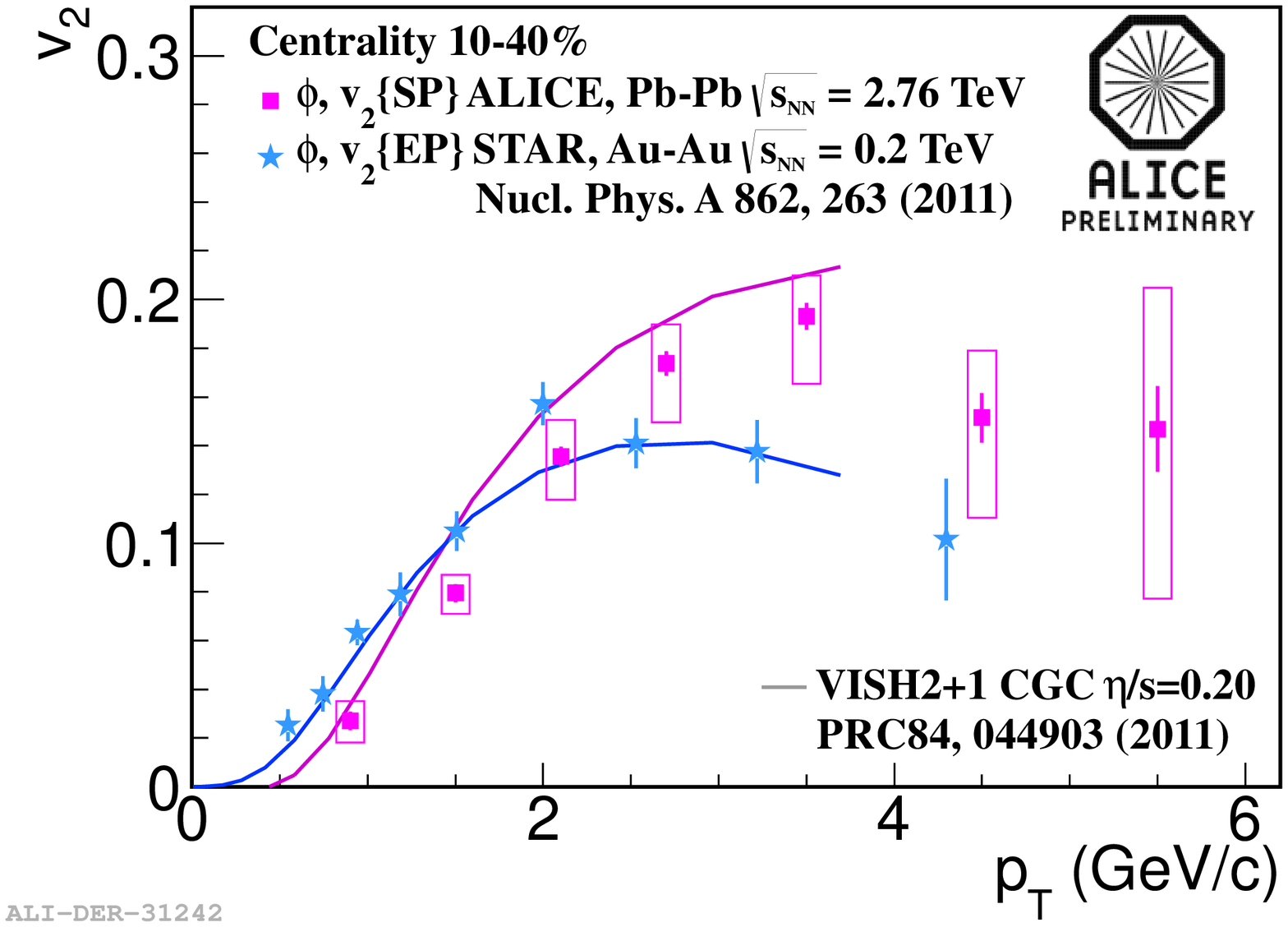}
\end{center}
\vspace{-0.8cm}
\caption{Identified particle $v_2(p_{\rm T})$ measured by ALICE compared to PHENIX \cite{bib:phenix} for 10-20\% centrality (left) and compared to STAR \cite{bib:star} results at RHIC for 10-40\% (right) centrality.}
\label{fig:v2rhic}
\end{figure}

\begin{figure}[hbp]
\begin{center}
 \includegraphics[width=0.49\textwidth]{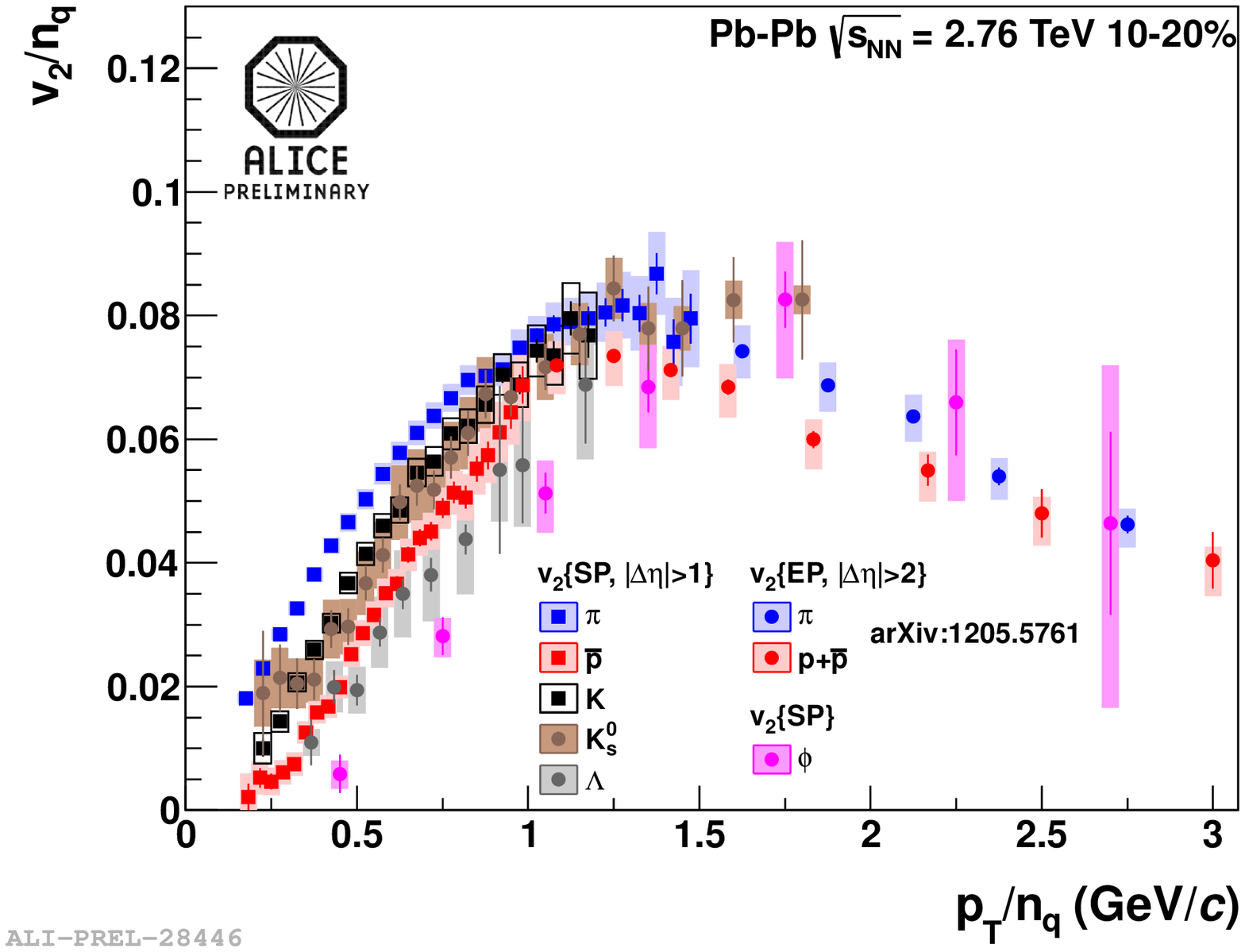}
 \includegraphics[width=0.49\textwidth]{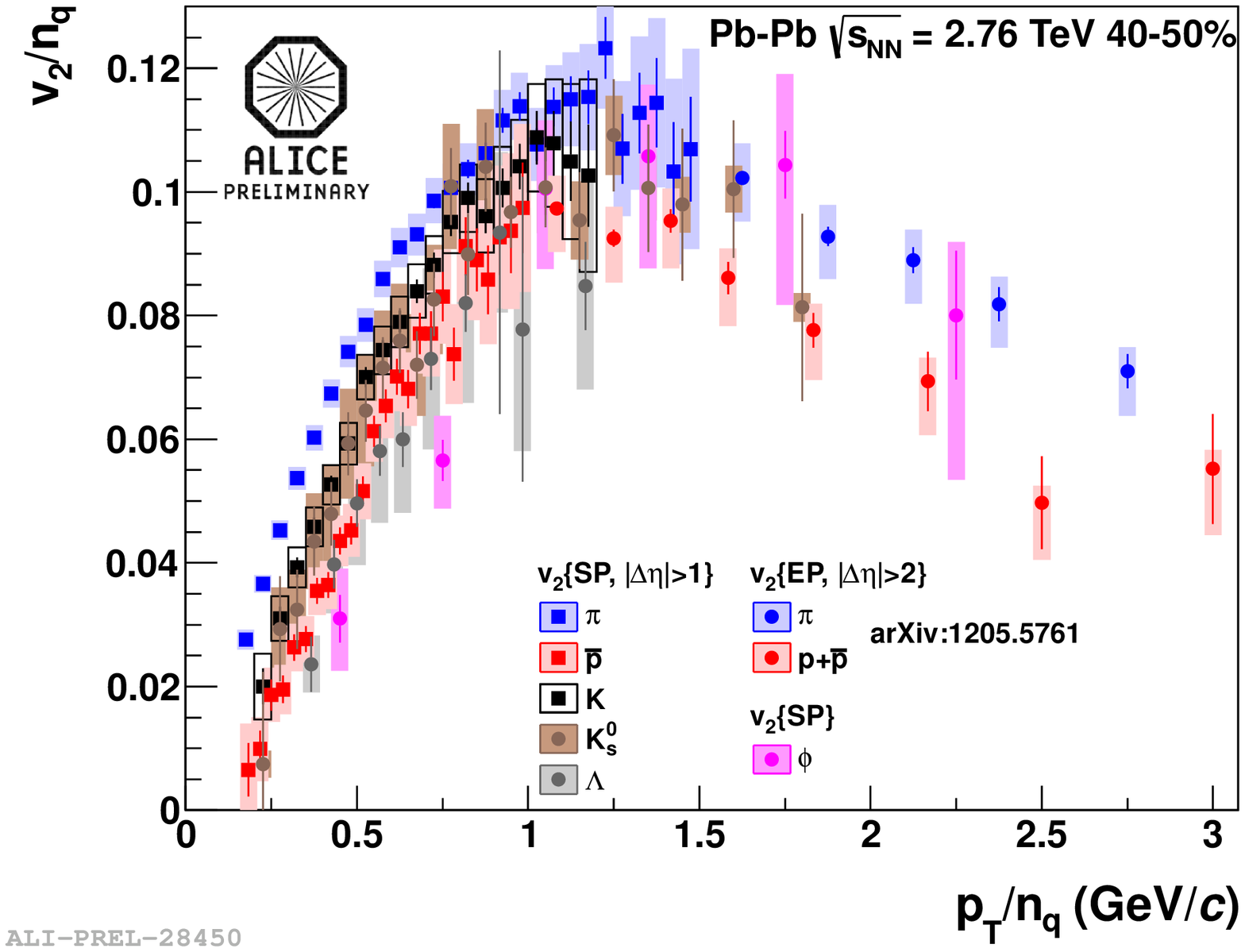}
\end{center}
\vspace{-0.8cm}
\caption{Identified particle $v_2(p_{\rm T})$ scaling with the constituent number of quarks, $n_q$, vs. $p_{\rm T}/n_q$ for 10-20\% (left) and 40-50\% (right) centrality classes.}
\label{fig:v2pt}
\end{figure}

\begin{figure}[hbp]
\begin{center}
 \includegraphics[width=0.49\textwidth]{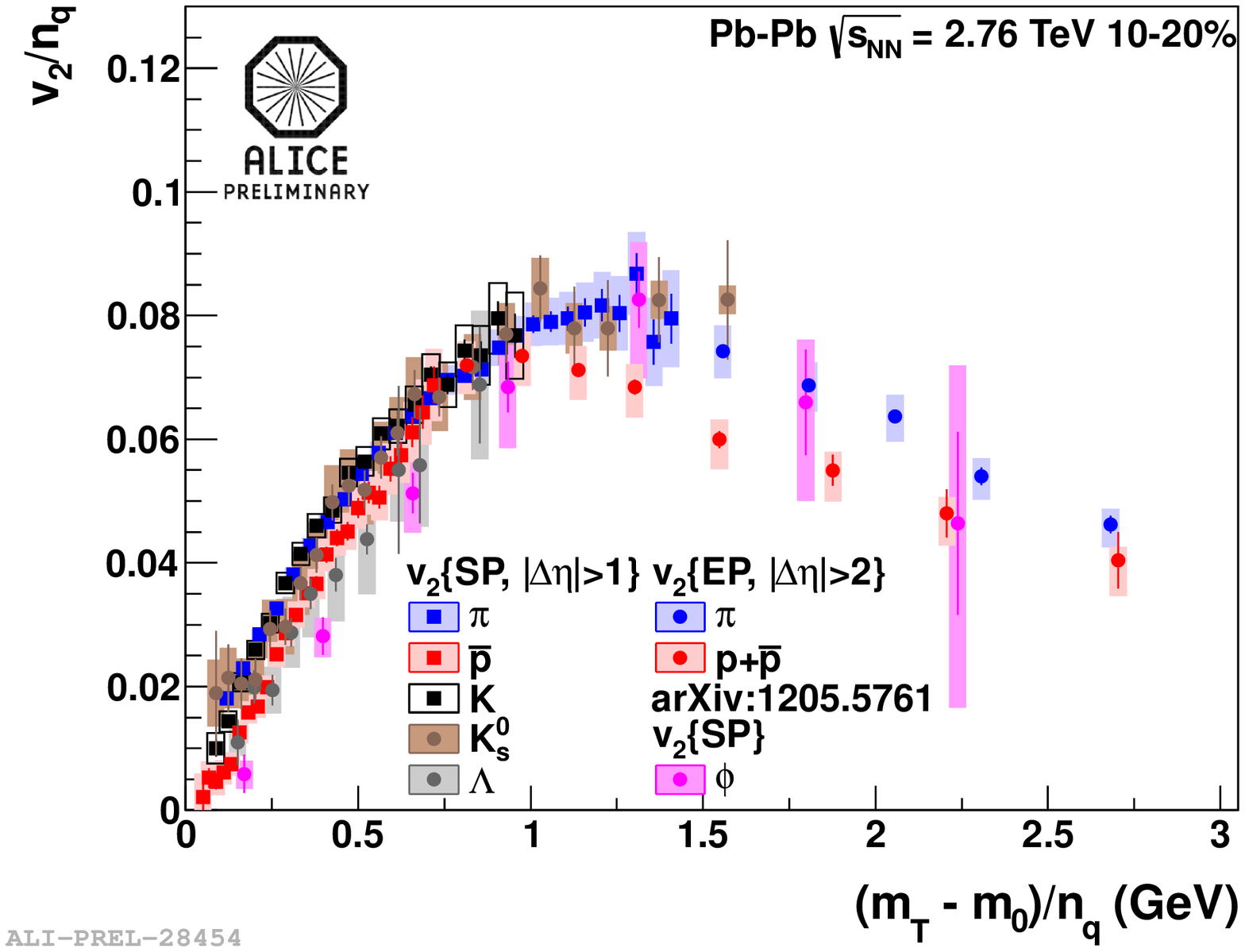}
 \includegraphics[width=0.49\textwidth]{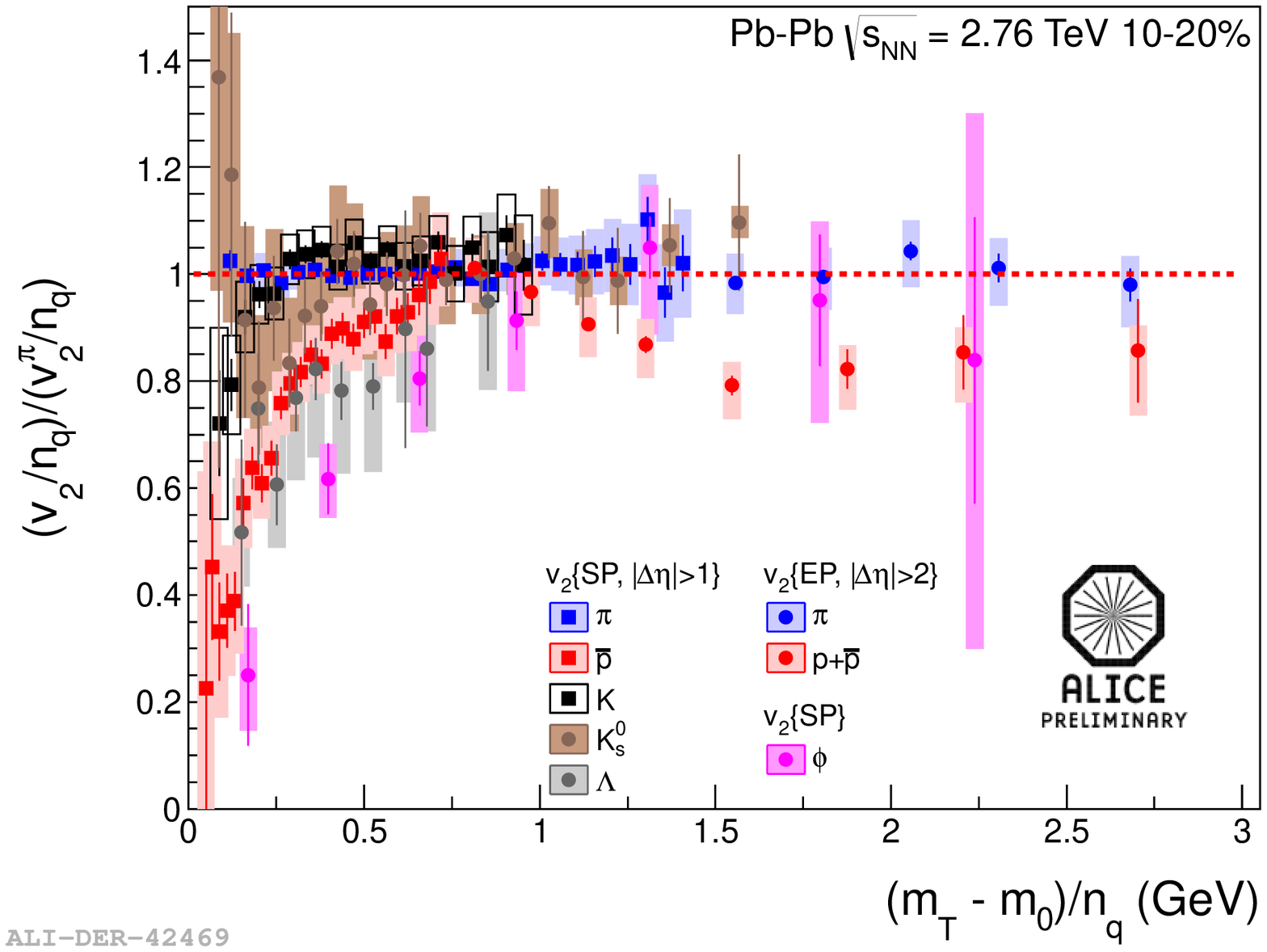}
\end{center}
\vspace{-0.8cm}
\caption{Identified particle $v_2(p_{\rm T})$ scaling with the constituent number of quarks, $n_q$, vs. transverse kinetic energy, $KE_{\rm T}$, per $n_q$ for 10-20\% (left). The ratio with respect pion scaled $v_2$ is also reported (right).}
\label{fig:v2mt}
\end{figure}

The measured $v_2$ of pions and kaons at LHC energy is slightly above the RHIC results, while anti-proton and ${\rm \phi}$-meson $v_2$
shows effect of the stronger radial flow at LHC energy, as expected from
hydrodynamical model calculations \cite{bib:hydroPhi}.

Figures \ref{fig:v2pt} and \ref{fig:v2mt} present the elliptic flow of identified particles
scaled with the constituent number of quarks, $n_q$
(2 for mesons and 3 for baryons)
vs. $p_{\rm T}$ and transverse kinetic energy, $KE_{\rm T} = \sqrt{m^2 + p_{\rm T}^2}$ per $n_q$.
In the intermediate $p_{\rm T}$ region (3-6 GeV/$c$) $v_2/n_q$ vs. $p_{\rm T}/n_q$ scaling serves as a test
of the hadron production via quark coalescence.
Identified particle flow at the LHC shows approximate, within 20\%, scaling of $v_2$ vs. $p_T$ with $n_q$ at $p_T \sim 1.2~{\rm GeV/}c$.
Shown in Fig.~\ref{fig:v2mt} the pion and kaon $v_2/n_q$ vs. $KE_{\rm T}/n_q$ follow the same scaled
$v_2$ while heavier particles, including ${\rm \phi}$-meson, deviate notably from meson flow
over the whole $p_{\rm T}$ range and especially in the central collisions.

\begin{figure}[htbp]
\begin{center}
 \includegraphics[width=0.99\textwidth]{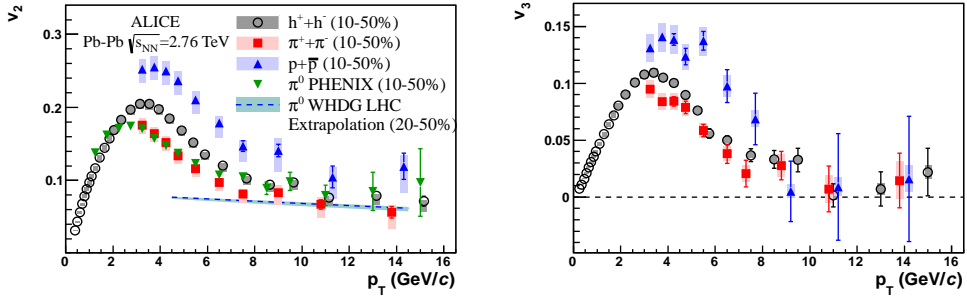}
\end{center}
\vspace{-0.8cm}
\caption{Unidentified charged hadrons, pion and (anti-)proton $v_2$ (left) and $v_3$ (right) vs. $p_{\rm T}$.
Pion $v_2(p_{\rm T})$ compared to that of neutral pion flow at RHIC \cite{bib:phenixHM}
 and WHDG model calculations \cite{bib:WHDG} for neutral pion extrapolated to the LHC collision energy.}
\label{fig:v2v3}
\end{figure}

Triangular flow of identified particles measured by ALICE shows a mass ordering similar to $v_2$ \cite{bib:QM11}.
Figure \ref{fig:v2v3} shows $v_2$ and $v_3$ at intermediate and high (above $p_{\rm T} \sim$ 6 GeV/$c$) transverse momenta
where hadron production from hard processes is expected to dominate.
Both $v_2$ and $v_3$ are finite up to 10 GeV/$c$ and
the proton $v_{2}$ $(v_{3})$ is larger than the pion flow up to 8 GeV/$c$.
The charged pion $v_2$ is similar to the ${\rm \pi^{0}}$ $v_2$ measured by PHENIX at
RHIC \cite{bib:phenixHM}.
\vspace{-0.8cm}
\section{Summary}
\vspace{-0.3cm}
$v_2(p_{\rm T})$ of identified particles is measured for Pb--Pb collisions at
$\sqrt{s_{\rm NN}}$ = 2.76 TeV by the ALICE detector.
Viscous hydrodynamic models describe
the main features of the elliptic flow of
all measured species.
Compared to the measurements at top RHIC energy, we observed a larger mass
splitting, mostly apparent in the proton and ${\rm \phi}$-meson flow.
Elliptic flow divided by the number of constituent quarks, $n_q$,
vs. $p_{\rm T}/n_q$ scales within 20\% for all identified species for  $p_T \sim 1.2~{\rm GeV/}c$.
$KE_{\rm T}/n_q$ scaling of $v_2$ is broken over the whole range of transverse momentum
in contrast to the observation at the top RHIC energy.
The measurement of $v_3(p_{\rm T})$ of proton and pion up to the high $p_{\rm T}$ = 16 GeV/$c$ is reported.
At large $p_{\rm T}$ the proton $v_2$ and $v_3$ are above pion flow at least up to $p_{\rm T}$ = 8 GeV/$c$.


\vspace{-0.3cm}
\begin{thebibliography}{00} 
\vspace{-0.3cm}
\bibitem{bib:v2} J.Y.Ollitrault, Phys. Rev. D 46, {\bf 229} (1992).
\bibitem{bib:v2phenom} S.Voloshin, Y.Zhang, Z. Phys. C {\bf 70}, 665 (1996).
\bibitem{bib:alice} ALICE Collaboration, arXiv:1205.5761 (2012).
\bibitem{bib:hydro} U. W. Heinz, C. Shen and H. Song, AIP Conf. Proc. {\bf 1441}, 766 (2012).
\bibitem{bib:phenix} PHENIX Collaboration, Phys. Rev. C {\bf 85}, 064914 (2012).
\bibitem{bib:star} STAR Collaboration, Nucl. Phys. A {\bf 862}, 263 (2011).
\bibitem{bib:hydroPhi} C. Shen, U. W. Heinz, P. Huovinen and H. Song, Phys. Rev. C {\bf 84}, 044903 (2011).
\bibitem{bib:QM11} M. Krzewicki [ALICE Collaboration], J. Phys. G {\bf 38}, 124047 (2011). 
\bibitem{bib:phenixHM} PHENIX Collaboration, Phys. Rev. Lett. {\bf 105}, 142301 (2010).
\bibitem{bib:WHDG} W. A. Horowitz and M. Gyulassy, J. Phys. G {\bf 38}, 124114 (2011).
\end{thebibliography}
\end{document}